\documentclass{raa}            
\usepackage{graphicx,times}             
\usepackage{natbib}
\usepackage{amssymb,amsmath}
\bibpunct{(}{)}{;}{a}{}{,}

\usepackage[a4paper=true,pagebackref=true]{hyperref}
\hypersetup{colorlinks = true, linkcolor = green, anchorcolor = red, citecolor = blue, filecolor = red, pagecolor = red, urlcolor = red}

\begin{document}

   \title{A photmetric study of the high-mass-ratio contact binary AV Puppis
}

   \volnopage{Vol.0 (20xx) No.0, 000--000}      
   \setcounter{page}{1}          

   \author{Quan-Wang Han
      \inst{1,2,3}
   \and Li-Fang Li
      \inst{1,2}
   \and Deng-Kai Jiang
      \inst{1,2}
   }

   \institute{Yunnan Observatories, Chinese Academy of Sciences, Kunming 650216, China; {\it qwhan@ynao.ac.cn}\\
        \and
             Key Laboratory of the Structure and Evolution of Celestial Objects, Chinese Academy of Sciences, Kunming 650216, China\\
        \and
             University of the Chinese Academy of Sciences, Beijing 100049, China\\
\vs\no
   {\small Received~~20xx month day; accepted~~20xx~~month day}}

\abstract{The multi-color photometric light curves for a contact binary AV Puppis  (AV Pup) in $VR_cI_c$ bandpasses are presented, and they are analyzed by using of the 2013 version of the Wilson-Devinney (W-D) code. The solutions suggest that AV Pup is a peculiar A-subtype W UMa contact binary with a high mass ratio ($q=m_2/m_1= 0.896$) and a fill-out factor ($f=10\%$). Combining with our newly determined times of minimum with those collected from literatures, the orbital period changes of this system are investigated. The $O-C$ analysis  shows that the orbital period of AV Pup is increasing at a rate of $\mathrm{d}P/\mathrm{d}t=4.83 \times 10^{-7} \ \mathrm{days \ yr^{-1}}$, which can be explained by mass transfer from the less massive component to the more massive one.
\keywords{techniques: photometric --- stars: magnetic field --- stars: individual: AV Pupis}
}

   \authorrunning{Q.-W. Han, L.-F. Li \& D.-K. Jiang}            
   \titlerunning{Photometry Study of AV Pup }  

   \maketitle

%
%
\section{Introduction}           
\label{sect:intro}
W Ursae Majoris contact binaries are the most common eclipsing systems around the solar system \citep{1948HarMo...7..249S}. They are short-period binary systems with both components filling their inner Roche lobes and sharing a common envelope.  The formation and evolutionary ending of contact binaries are still open questions in astrophysics. The most plausible scenario is that they are formed from detached binaries via angular momentum loss (AML) \citep{1982A&A...109...17V} or evolutionary expansion of the components \citep{1976ApJS...32..583W}. Model calculations suggest that this type of systems will ultimately evolve into contact binaries of extreme mass ratios or even into the fast-rotating single stars under the influence of AML \citep{2004MNRAS.355.1383L}.

AV Pup (GSC 05998-02010, $\alpha_{2000}$=$08^{h}24^{m}32.\!^{s}30$, $\delta_{2000}$=$-16^{\circ}24'11.\!''24$) was firstly reported   by \cite{1930AN....240..193H} as a variable star. \cite{1980AcA....30..501B} classified it as a semi-detached binary star with light curves of W UMa type. \cite{2005Ap&SS.300..289W} analyzed this system by using the V band photometric data of All Sky Automated Survey (ASAS) . He found this system is a contact binary with a high mass ratio of 0.80 and a low fill-out factor of 10\%.  After that, this system has been barely studied.

In this work,  we presented two years of photometric observations in $VR_cI_c$ bandpasses for AV Pup. We also obtained new photometric solutions and a period analysis  for this system. The article is organized as follows: in section 2, the new observations for AV Pup are shown; in section 3, a period analysis is conducted for AV Pup;  in section 4, the photometric solutions for the system are presented; at last, the summary and some discussions are given in section 5.

\section{Observations}
\label{sect:Obs}
\begin{figure}[!htbp]\centering
\includegraphics[width=10cm]{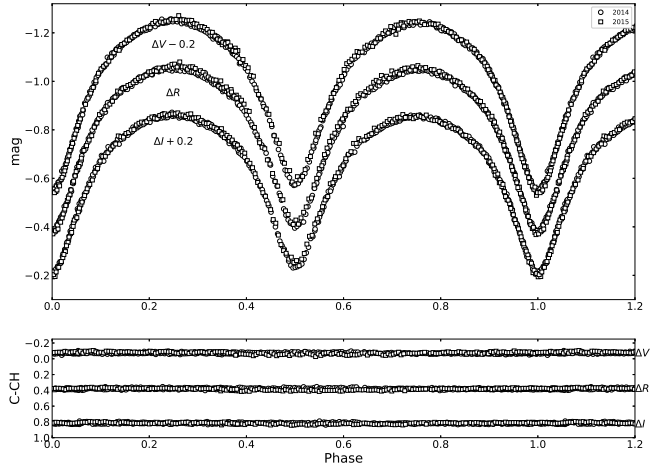}
\caption{The observed light curves in $VR_cI_c$ bandpasses for AV Pup. The bottom panel is the magnitude differences between comparison and check stars. Open circles and open squares represent the data in 2014 and 2015, respectively. }
\label{fig:observe}
\end{figure}

New $VR_cI_c$  photometric observations of  AV Pup were obtained in  2014 and 2015 by using the 1m Cassegrain reflector telescope at Yunnan Observatory. The telescope mounts an  Andor DW436 $2048 \times 2048$ CCD camera with an  effective field of view of 7.3  $\times$  7.3 $\rm{arcmin}^{2}$. The aperture photometry package in IRAF was used for the data reduction.  In our observations, the nearby stars TYC 5998-1820-1 and TYC 5998-2135-1 were employed as the comparison star and the check star, respectively. Their coordinates are listed in Table \ref{table:coo}. The new light curves observed in two recent observation seasons are shown in Figure \ref{fig:observe}. It is found in Figure \ref{fig:observe} that the two sets of light curves are quite similar, only have a slight difference between the phases 0.25 and 0.75. 
\begin{table}[!htbp]\centering
\caption{Coordinates of AV Pup, the comparison star and the check star.}\label{table:coo}
\begin{tabular}{lcccc}\hline\hline
Targets& Name & $\alpha_{J2000}$ & $\delta_{J2000}$ & $V_{\mathrm{mag}}$ \\\hline
Variable&AV Pup                  &08 24 32.30        &-16 24 11.24    & 10.68  \\
The comparison&  TYC 5998-1820-1  &08 24 15.30        &-16 23 13.43  & 11.56  \\
The check& TYC 5998-2135-1         &08 24 22.63        &-16 25 15.75 & 11.53  \\
\hline
\end{tabular}
\end{table}

Six new times of minimum were determined by using the K-W method \citep{1956BAN....12..327K} based on our observations. Then we took the average value of three bandpasses as one minimum. They are listed in Table \ref{table:newmini}.

\begin{table}[!htbp]\centering
\caption{New times of minimum for AV Pup}\label{table:newmini}
\begin{tabular}{cccc}\hline\hline
HJD& error &p/s&fiter\\\hline
2456715.2366 & 0.0001       & p & $VR_cI_c$    \\
2456716.1067 & 0.0001       & p & $VR_cI_c$    \\
2456717.1950 & 0.0001       & s & $VR_cI_c$    \\
2457094.1343 & 0.0002       & p & $VR_cI_c$    \\
2457096.0923 & 0.0002       & s & $VR_cI_c$    \\
2457098.0494 & 0.0001       & p & $VR_cI_c$    \\
\hline
\end{tabular}
\end{table}

\section{Orbital period analysis}
\label{sect:orbital}
Besides the new times of minimum derived by us from our observations, we also collected other times of minimum from the $O-C$ Gateway database\footnote{http://var.astro.cz/ocgate/}, AAVSO\footnote{https://www.aavso.org/}  and literatures. We only used the CCD data (listed in Table \ref{table:minima}) with a relatively high accuracy to investigate the changes in the orbital period of AV Pup, since the visual observations are too dispersive (in $O-C$ diagram) to have a significant help for the analysis of period changes of this object. \cite{2005Ap&SS.300..289W} assigned the orbital period of AV Pup as 0.435010 d, which is different from 0.556339 d obtained by \cite{1980AcA....30..501B}. We find the former one seems to be more reliable based on our new observations.  Based on the orbital period 0.435010 d and a primary minimum HJD 2456715.2366,  we can give the following linear ephemeris:
\begin{equation}
\mathrm{Min.I}= \mathrm{HJD}\ 2456715.2366+ 0.435010\times E.
\label{eq:ephoc}
\end{equation}

\begin{table*}[!htbp]\centering
\caption{All CCD times of minimum for AV Pup.}
\label{table:minima}
\begin{tabular}{ccccccl}

\hline\hline
HJD&Error&Epoch& O-C &Min&Filter&References\\
(2400000+) &&&&&\\\hline
52002.5424 & - & -10833.5 & -0.0134 & II & CCD & AAVSO  \\ 
52237.2300 & - & -10294.0 & -0.0137 & I & $R_c$ &  \cite{vsolj-39}   \\ 
52655.0563 & - & -9333.5 & -0.0145 & II & $R_c$ &  \cite{vsolj-42}   \\ 
53020.0299 & - & -8494.5 & -0.0143 & II & $R_c$ &  \cite{vsolj-43}   \\ 
53021.7705 & - & -8490.5 & -0.0137 & II & CCD & AAVSO   \\ 
53035.6907 & - & -8458.5 & -0.0138 & II & CCD & AAVSO   \\ 
53043.0860 & - & -8441.5 & -0.0137 & II & $R_c$ &  \cite{vsolj-43}   \\ 
53074.6232 & - & -8369.0 & -0.0147 & I & CCD &  AAVSO   \\ 
53109.6840 & - & -8288.5 & 0.0278 & II & CCD &  AAVSO   \\ 
53403.0552 & - & -7614.0 & -0.0153 & I & $R_c$ &   \cite{vsolj-44}  \\ 
53405.0126 & - & -7609.5 & -0.0154 & II & $R_c$ &  \cite{vsolj-44}   \\ 
53426.9828 & - & -7559.0 & -0.0132 & I & $V$ &    \cite{vsolj-44} \\ 
53743.2333 & - & -6832.0 & -0.0150 & I & $V$ &    \cite{vsolj-45} \\ 
53761.0689 & - & -6791.0 & -0.0148 & I & $V$ &    \cite{vsolj-45} \\ 
54119.0815 & - & -5968.0 & -0.0154 & I & $V$ &    \cite{vsolj-46} \\ 
54126.0425 & - & -5952.0 & -0.0146 & I & $V$ &    \cite{vsolj-46} \\ 
54127.7825 & 0.0003 & -5948.0 & -0.0146 & I & CCD &  \cite{2008JAVSO..36..186S}   \\ 
54526.6873 & 0.0001 & -5031.0 & -0.0140 & I & CCD &  \cite{2008JAVSO..36..186S}   \\ 
54526.6874 & 0.0002 & -5031.0 & -0.0139 & I & CCD &  \cite{2008JAVSO..36..186S}   \\
54545.6101 & 0.0002 & -4987.5 & -0.0141 & II & CCD & \cite{2008JAVSO..36..186S}    \\ 
54548.0023 & - & -4982.0 & -0.0145 & I & $V$ &    \cite{vsolj-48} \\ 
54877.7380 & 0.0004 & -4224.0 & -0.0164 & I & CCD &  \cite{2009JAVSO..37...44S}   \\ 
54901.6674 & 0.0002 & -4169.0 & -0.0125 & I & CCD &  \cite{2010JAVSO..38...85S}   \\ 
55175.9443 & 0.0005 & -3538.5 & -0.0094 & II & CCD & \cite{2010JAVSO..38..183S}    \\ 
55232.7119 & 0.0003 & -3408.0 & -0.0106 & I & CCD &  \cite{2010JAVSO..38..183S}   \\ 
55271.6457 & 0.0001 & -3318.5 & -0.0102 & II & CCD & \cite{2011JAVSO..39...94S}    \\ 
55612.6958 & 0.0002 & -2534.5 & -0.0080 & II & $V$ &   \cite{2011JAVSO..39..177S}  \\ 
55622.7003 & 0.0002 & -2511.5 & -0.0087 & II & $V$ &   \cite{2011JAVSO..39..177S}  \\ 
55630.7481 & 0.0002 & -2493.0 & -0.0086 & I & $V$ &    \cite{2011IBVS.5992....1D} \\ 
56290.8767 & 0.0015 & -975.5 & -0.0076 & II & $V$ &    \cite{2013IBVS.6042....1D} \\ 
56737.6404 & 0.0001 & 51.5 & 0.0008 & II & $V$ &     \cite{2014JAVSO..42..426S}\\ 
56715.2366 & 0.0001 & 0.0 & 0.0000 & I & $VR_cI_c$ &     This paper \\ 
56716.1067 & 0.0001 & 2.0 & 0.0001 & I & $VR_cI_c$ &     This paper \\ 
56717.1950 & 0.0001 & 4.5 & 0.0009 & II & $VR_cI_c$ &    This paper  \\ 
57034.5430 & 0.0040 & 734.0 & 0.0091 & I & CCD &    \cite{2015OEJV..172....1P} \\ 
57094.1343 & 0.0002 & 871.0 & 0.0040 & I & $VR_cI_c$ &   This paper   \\ 
57096.0923 & 0.0002 & 875.5 & 0.0044 & II & $VR_cI_c$ &  This paper    \\ 
57098.0494 & 0.0001 & 880.0 & 0.0040 & I & $VR_cI_c$ &   This paper   \\
58133.1690 & -      & 3259.5 & 0.0044 & II & $BVI_c$ &  \cite{vsolj-66}    \\ 
58135.7783 & 0.0001 & 3265.5 & 0.0040 & II & $V$      &   AAVSO    \\
58203.6405 & 0.0001 & 3421.5 & 0.0040 & II & $V$      &   AAVSO    \\\\
\hline
\end{tabular}\\
\end{table*}

\begin{figure}[!htbp]\centering
\includegraphics[width=10cm]{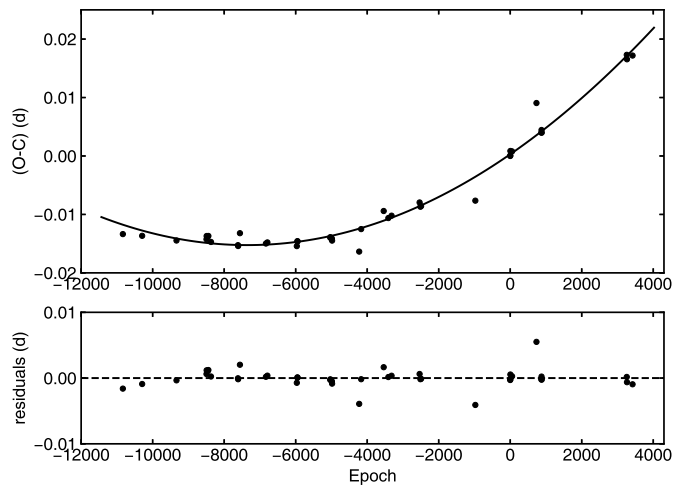}
\caption{$O-C$ diagram  of AV Pup. The points are the $O-C$ values calculated with Equation \ref{eq:ephoc}. The solid line represents a least-square solution. The lower panel shows the corresponding residuals.}
\label{fig:ephem}
\end{figure}

The $O-C$ values are calculated based on Equation \ref{eq:ephoc} and listed in the fourth column of Table \ref{table:minima}. As seen in Table \ref{table:minima}, the minimum 2453109.6840 evidently deviates from the other minima, so we neglected this value. The $O-C$ values are shown in Figure \ref{fig:ephem}. A clear parabola track can be seen from this Figure. We use a  least-square solution to fit all available times of minima and get the following quadratic ephemeris:
\begin{equation}
\mathrm{Min.I}= \mathrm{HJD}\ 2456715.2369(3)+ 0.4350142(1) \times E +2.88(16) \times 10^{-10} \times E^2.
\label{eq:ephquad}
\end{equation}
The period of AV Pup shows a secular increase. The increasing rate is derived as $\mathrm{d}P/\mathrm{d}t=4.83 \times 10^{-7} \ \mathrm{days \ yr^{-1}}$ based on Equation \ref{eq:ephquad}. Since the CCD times of minimum only last less than 20 years, there is no clear cyclic variation showing in the $O-C$ diagram.

\section{Light curve solution}
We analyzed  our new light curves  based on the 2013 version of Wilson-Divinney (W-D) code \citep{1971ApJ...166..605W, 1979ApJ...234.1054W, 1990ApJ...356..613W}. \cite{2005Ap&SS.300..289W} assigned the surface temperature of AV Pup as 6255 K, which is corresponding to a spectral type of F8 in the General Catalog of Variable Stars.  We took this value as the effective temperature of star 1 (eclipsed at phase 0.0) in our calculation. This effective temperature means that AV Pup should have a common convective envelope, so the bolometric albedos and gravity-darkening coefficients were set as  $A_1=A_2=0.5$ \citep{1969AcA....19..245R} and  $g_1=g_2=0.32$ \citep{1967ZA.....65...89L}, respectively.  We used the logarithmic law format of limb-darkening coefficients \citep{1993AJ....106.2096V}, which were computed internally by the DC program of W-D code.  

\begin{figure}[!htbp]\centering
\includegraphics[width=10cm]{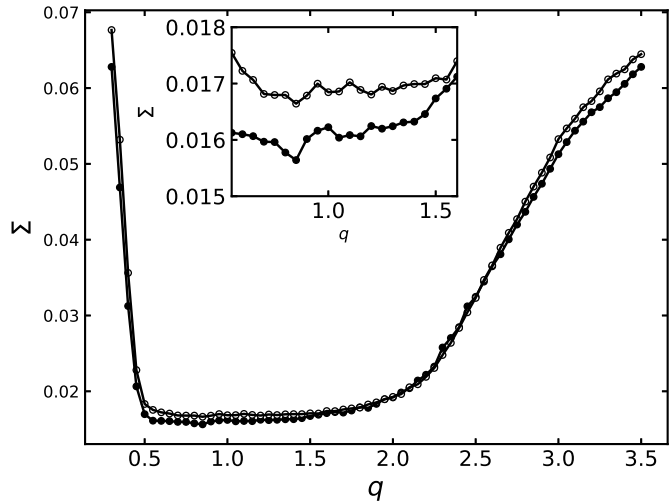}
\caption{ The relation between $\Sigma$ (the mean residuals for input data) and the mass ratio $q$ of AV Pup. The filled circles and open circles represent for data of 2014 and 2015, respectively. }
\label{fig:qsearch}
\end{figure}

Due to  the lack of spectroscopic mass ratio, a q-search procedure was performed to determine the initial mass ratio of AV Pup. A series of fixed mass ratios ranged from 0.1 to 5.0 in a step of 0.05 were adopted. In the q-search procedure, the adjustable parameters in our calculation were: the orbital inclination ($i$), the effective temperature of star 2 ($T_2$ ), the surface potential ($\Omega_1$ and $\Omega_2$) and the bandpass luminosity of star 1 ($L_1$). Due to the diffusion of semi-detached configuration and contact configuration, we ran DC subroutine of W-D code beginning with Mode 2 (a detached configuration) for each fixed $q$. However, the program quickly converged  to Mode 3 (a contact configuration) at last, which means AV Pup should be a contact binary, as derived by \cite{2005Ap&SS.300..289W}. For a clear view, Figure \ref{fig:qsearch} only shows the relation between $\Sigma$ (the mean residuals for input data)  and $q$ in the range of $q=0.3$ to 3.5. As seen from Figure \ref{fig:qsearch}, the results of two q-search procedures based on the light curves observed in 2014 and 2015 are quite similar. They have a flat pattern from  $q=0.5$ to $q=2.0$ and with a minimum at  $q= 0.85$. We took  $q= 0.85$  as an initial mass ratio and assigned $q$ to be an adjustable parameter for the later calculation.

\begin{figure}[!htbp]\centering
\includegraphics[width=10cm]{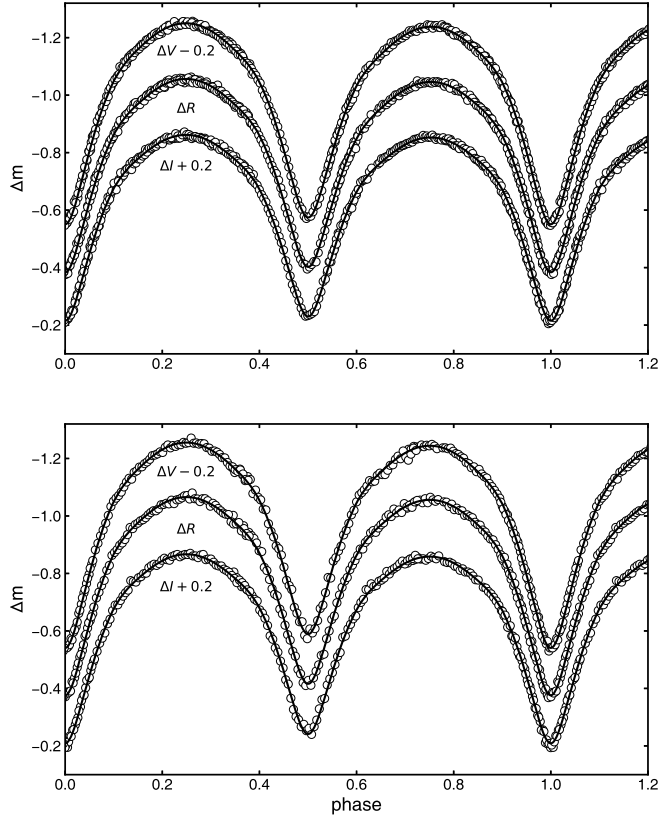}
\caption{Comparison of observed (open circles) and computed (black lines) light curves for AV Pup. The upper and bottom panel show the comparison of 2014 and 2015, respectively.}
\label{fig:theorylc}
\end{figure}

As seen from Figure \ref{fig:observe}, a slight O'Connell effect is shown in the light curves of AV Pup, so we attempted to model the light curves with spots. During the calculation, a third light was also taken into account. After some attempts, we finally got the best convergent solutions with a cool spot located on star 1 for the data of two years.  The results  are given in Table \ref{table:solu} and the comparison between observed and computed light curves is shown in Figure \ref{fig:theorylc}.

\begin{table}[!htbp]\centering
\caption{Photometric solutions of AV Pup.}
\label{table:solu}
\begin{tabular}{ccc}\hline\hline
Parameters & 2014 &2015 \\\hline
$g_1=g_2$         & 0.32(fixed)  & 0.32(fixed)  \\
$A_1=A_2$         & 0.5(fixed)   & 0.5(fixed) \\
$T_1(K)$          & 6255(fixed)  & 6255(fixed) \\
$T_2(K)$          & 6145 $\pm$ 7 & 6150 $\pm$ 9\\
$q(M_2/M_1)$      & 0.896 $\pm$ 0.005 & 0.896 $\pm$ 0.003 \\
$\Omega_{1}=\Omega_{2}$   & 3.525 $\pm$ 0.008         & 3.529 $\pm$ 0.005\\
$i(deg)$                 & 81.222 $\pm$ 0.101  & 80.845 $\pm$ 0.109 \\

$L_{1}/(L_{1}+L_{2}+L_{3})_{V}$ & 0.516 $\pm$ 0.003 & 0.530 $\pm$ 0.008 \\
$L_{1}/(L_{1}+L_{2}+L_{3})_{R_c}$ & 0.509 $\pm$ 0.003 & 0.524 $\pm$ 0.007 \\
$L_{1}/(L_{1}+L_{2}+L_{3})_{I_c}$ & 0.500 $\pm$ 0.003 & 0.512 $\pm$ 0.006 \\
$L_{3}/(L_{1}+L_{2}+L_{3})_{V}$ & 0.052 $\pm$ 0.003 & 0.025 $\pm$ 0.010 \\
$L_{3}/(L_{1}+L_{2}+L_{3})_{R_c}$ & 0.060 $\pm$ 0.003 & 0.031 $\pm$ 0.009 \\
$L_{3}/(L_{1}+L_{2}+L_{3})_{I_c}$ & 0.071 $\pm$ 0.003 & 0.048 $\pm$ 0.008 \\
$r_1(pole)$   & 0.3726 $\pm$  0.0005    & 0.3720 $\pm$  0.0004\\
$r_1(side)$   & 0.3932 $\pm$  0.0005    & 0.3924 $\pm$  0.0005\\
$r_1(back)$   & 0.4273 $\pm$  0.0005    & 0.4263 $\pm$  0.0006\\
$r_2(pole)$   & 0.3536 $\pm$  0.0018    & 0.3538 $\pm$  0.0011\\
$r_2(side)$   & 0.3721 $\pm$  0.0023    & 0.3723 $\pm$  0.0014\\
$r_2(back)$   & 0.4070 $\pm$  0.0036    & 0.4072 $\pm$  0.0022\\
$f(\%)$   & 10.9  $\pm$  1.6 & 10.2  $\pm$  1.1\\
Spot parameters: &\\
Latitude(deg)   &  24.4    &  29.6   \\
Longitude(deg)  &  144.0    &  97.4   \\
Radius(deg)     &  31.0   &  16.9   \\
T/T$_1$           &  0.92    &  0.89   \\

\hline
\end{tabular}
\end{table}
Our solutions suggest that AV Pup is an A-subtype contact binary. It has a high mass ratio of 0.896 and an inclination around 81$^\circ$. The system is a shallow contact system with a fill-out factor around 10\% , which coincides with the result of \cite{2005Ap&SS.300..289W}. The slight difference between two years light curves can be explained by the spot activity. Because there is no radial velocity curves observed for this system, we can not obtain the precise absolute parameters for the components of AV Pup. Therefore, we used the  mass-temperature relation \citep{1988BAICz..39..329H} to derive the mass of the primary star, the primary's mass can be estimated as $M_1= 1.27~M_{\odot}$, then the mass of secondary star was determined to be $M_2=1.14\pm0.01~M_{\odot}$ based on our photometric solution. The radii and luminosity for two components can also be obtained as: $R_1=1.29\pm0.01~R_{\odot}$, $L_1=2.29\pm0.02~L_{\odot}$,  $R_2=1.23\pm0.02~R_{\odot}$ and $L_2=1.94\pm0.10~L_{\odot}$.

\section{Summary and discussion}

In this paper, we presented two years of CCD photometric observations for AV Pup. We derived the photometric solutions for this system. Two solutions show that AV Pup is an A-subtype contact binary with a high mass ratio of $q=m_2/m_1= 0.896$. The system is a shallow contact binary with a low fill-out factor around 10\%. Two years of observations show a slight difference, which can be explained by the spot variation.  Contact binaries tend to have a low mass ratio \citep{2001AJ....122.1007R}, so only a few of contact binaries with high mass ratio were discovered, such as WZ And \citep[$q=1.0$,][]{2006NewA...11..339Z},  HT Vir \citep[$q=0.9$,][]{2014IBVS.6121....1B}  and so on. Meanwhile, the A-subtype contact binaries have a relatively low mass ratio and a relatively high contact degree in general \citep{1988MNRAS.231..341H,2009MNRAS.396.2176J}. However, a few A-subtype contact binaries were found to have a high mass ratio and a shallow common envelope (listed in Table \ref{table:atype}). AV Pup seems to have the same characteristics as these peculiar contact binaries.
 
\begin{table}[!htbp]\centering
\caption{A-subtype contact binaries with a high mass ratio}
\label{table:atype}
\begin{tabular}{cccccl}\hline\hline
Name &$T_1\ (K)$ & $T_2\ (K)$ & $q$  & $f\ (\%)$  & References\\\hline
OO Aql   & 5700   & 5472 & 0.844  & 21.4 & \cite{2013AJ....145..127I}  \\
V2150 Cyg & 8000   & 7920 & 0.802  & 19.0 & \cite{2003AA...412..465K}  \\
V1101 Her & 5920   & 5690 & 0.800  & 14.2 & \cite{2017AJ....154..260P}  \\
DZ Lyn     &   6860  &  5068 & 0.886  & 18.0 & \cite{2009RAA.....9.1270M}  \\
AU Ser    & 5495   & 5153 & 0.710  & 19.8 & \cite{2005NewA...10..653G}  \\
\hline
\end{tabular}
\end{table}

We also conducted the first period change analysis for AV Pup. A least-square fitting  shows that  the orbital period of  AV Pup is suffering  a secular increase at a rate of $\mathrm{d}P/\mathrm{d}t=4.83 \times 10^{-7} \ \mathrm{days \ yr}^{-1}$, which may be caused by the mass transfer from the less massive component to  the more massive one. Assuming the mass transfer is conservative, we can get the mass transfer rate from the formula derived from the Kepler law:
\begin{equation}
\frac{\mathrm{d}M_2}{\mathrm{d}t} = \dot M_{2}= \frac{{\dot P{M_1}{M_2}}}{{3P({M_2} - {M_1})}}.
\label{eq:mt}
\end{equation}
The estimated mass transfer rate is $\mathrm{d}M_2/\mathrm{d}t=-4.12 \times 10^{-6} \ M_{\odot} \ \rm{yr}^{-1}$. This mass transfer rate seems quite high for contact binaries, but it coincides with the mass transfer properties of contact binaries in the Kepler eclipsing binary catalogue \citep{2018PASJ...70...90K}.  The time scale of mass transfer for the donor star can be estimated as $\tau_{_{\rm MT}}=2.77 \times 10^{5}$ yr, which is much shorter than the thermal time scale $\tau_{\rm th} \sim(GM^{2})/(RL)\sim1.70 \times 10^7$ yr \citep{1971ARA&A...9..183P}. This suggests that the donor star can not maintain its thermal equilibrium. This system might be in the phase evolving from a contact configuration to a semi-detached one of the  thermal relaxation oscillation \citep{1976ApJ...205..208L, 1976ApJ...205..217F, 2004MNRAS.351..137L, 2005MNRAS.360..272L, 2008MNRAS.387...97L}. In the same way, the period increase may be part of a long-period cyclic variation caused by a third body, which is reflected by the third light showing in the photometric solutions. By the way, we should notice that these results were achieved according to the  absolute parameters only derived from the photometric observations, so spectroscopic observations are urgently needed.

\begin{acknowledgements}
This work was partly supported by the Chinese Natural Science Foundations (Nos. 11773065,  11573061, 11573062, 11390374 and 11661161016), and the Yunnan Natural Science Foundation (Grant No. 2015FB190). New CCD photometric observations were obtained with the 1.0 m telescope at Yunnan observatory.
\end{acknowledgements}

\begin{thebibliography}{50}
\providecommand\natexlab[1]{#1}

\bibitem[{Bensch} {et~al.}(2014)]{2014IBVS.6121....1B}
{Bensch}, K., {Dimitrov}, W., {Zywucka}, N., {et~al.} 2014, IBVS, 6121

\bibitem[{Brancewicz} \& {Dworak}(1980)]{1980AcA....30..501B}
{Brancewicz}, H.~K., \& {Dworak}, T.~Z. 1980, \actaa, 30, 501

\bibitem[{Diethelm}(2011)]{2011IBVS.5992....1D}
{Diethelm}, R. 2011, IBVS, 5992

\bibitem[{Diethelm}(2013)]{2013IBVS.6042....1D}
{Diethelm}, R. 2013, IBVS, 6042

\bibitem[{Flannery}(1976)]{1976ApJ...205..217F}
{Flannery}, B.~P. 1976, \apj, 205, 217

\bibitem[{G{\"u}rol}(2005)]{2005NewA...10..653G}
{G{\"u}rol}, B. 2005, \na, 10, 653

\bibitem[{Harmanec}(1988)]{1988BAICz..39..329H}
{Harmanec}, P. 1988, Bulletin of the Astronomical Institutes of Czechoslovakia,
  39, 329

\bibitem[{Hilditch} {et~al.}(1988)]{1988MNRAS.231..341H}
{Hilditch}, R.~W., {King}, D.~J., \& {McFarlane}, T.~M. 1988, \mnras, 231, 341

\bibitem[{Hoffmeister}(1930)]{1930AN....240..193H}
{Hoffmeister}, C. 1930, Astronomische Nachrichten, 240, 193

\bibitem[{{\.I}{\c c}li} {et~al.}(2013)]{2013AJ....145..127I}
{{\.I}{\c c}li}, T., {Ko{\c c}ak}, D., {Boz}, G.~{\c C}., \& {Yakut}, K. 2013,
  \aj, 145, 127

\bibitem[{Jiang} {et~al.}(2009)]{2009MNRAS.396.2176J}
{Jiang}, D., {Han}, Z., {Jiang}, T., \& {Li}, L. 2009, \mnras, 396, 2176

\bibitem[{Kouzuma}(2018)]{2018PASJ...70...90K}
{Kouzuma}, S. 2018, \pasj, 70, 90

\bibitem[{Kreiner} {et~al.}(2003)]{2003AA...412..465K}
{Kreiner}, J.~M., {Rucinski}, S.~M., {Zola}, S., {et~al.} 2003, \aap, 412, 465

\bibitem[{Kwee} \& {van Woerden}(1956)]{1956BAN....12..327K}
{Kwee}, K.~K., \& {van Woerden}, H. 1956, \bain, 12, 327

\bibitem[{Li} {et~al.}(2004{\natexlab{a}})]{2004MNRAS.351..137L}
{Li}, L., {Han}, Z., \& {Zhang}, F. 2004{\natexlab{a}}, \mnras, 351, 137

\bibitem[{Li} {et~al.}(2004{\natexlab{b}})]{2004MNRAS.355.1383L}
{Li}, L., {Han}, Z., \& {Zhang}, F. 2004{\natexlab{b}}, \mnras, 355, 1383

\bibitem[{Li} {et~al.}(2005)]{2005MNRAS.360..272L}
{Li}, L., {Han}, Z., \& {Zhang}, F. 2005, \mnras, 360, 272

\bibitem[{Li} {et~al.}(2008)]{2008MNRAS.387...97L}
{Li}, L., {Zhang}, F., {Han}, Z., {Jiang}, D., \& {Jiang}, T. 2008, \mnras,
  387, 97

\bibitem[{Lucy}(1967)]{1967ZA.....65...89L}
{Lucy}, L.~B. 1967, \zap, 65, 89

\bibitem[{Lucy}(1976)]{1976ApJ...205..208L}
{Lucy}, L.~B. 1976, \apj, 205, 208

\bibitem[{Martignoni} {et~al.}(2009)]{2009RAA.....9.1270M}
{Martignoni}, M., {Acerbi}, F., \& {Barani}, C. 2009, \raa, 9, 1270

\bibitem[{Nagai}(2001)]{vsolj-39}
{Nagai}, K. 2001, Variable Star Bulletin, 39, 5

\bibitem[{Nagai}(2004)]{vsolj-42}
{Nagai}, K. 2004, Variable Star Bulletin, 42, 1

\bibitem[{Nagai}(2005)]{vsolj-43}
{Nagai}, K. 2005, Variable Star Bulletin, 43, 1

\bibitem[{Nagai}(2006)]{vsolj-44}
{Nagai}, K. 2006, Variable Star Bulletin, 44, 1

\bibitem[{Nagai}(2007)]{vsolj-45}
{Nagai}, K. 2007, Variable Star Bulletin, 45, 1

\bibitem[{Nagai}(2008)]{vsolj-46}
{Nagai}, K. 2008, Variable Star Bulletin, 42, 1

\bibitem[{Nagai}(2009)]{vsolj-48}
{Nagai}, K. 2009, Variable Star Bulletin, 48, 1

\bibitem[{Nagai}(2019)]{vsolj-66}
{Nagai}, K. 2019, Variable Star Bulletin, 66, 1

\bibitem[{Paczy{\'n}ski}(1971)]{1971ARA&A...9..183P}
{Paczy{\'n}ski}, B. 1971, \araa, 9, 183

\bibitem[{Paschke}(2015)]{2015OEJV..172....1P}
{Paschke}, A. 2015, Open European Journal on Variable Stars, 172, 1

\bibitem[{Pi} {et~al.}(2017)]{2017AJ....154..260P}
{Pi}, Q.~F., {Zhang}, L.~Y., {Bi}, S.~L., {et~al.} 2017, \aj, 154, 260

\bibitem[{Ruci{\'n}ski}(1969)]{1969AcA....19..245R}
{Ruci{\'n}ski}, S.~M. 1969, \actaa, 19, 245

\bibitem[{Rucinski}(2001)]{2001AJ....122.1007R}
{Rucinski}, S.~M. 2001, \aj, 122, 1007

\bibitem[{Samolyk}(2008)]{2008JAVSO..36..186S}
{Samolyk}, G. 2008, JAAVSO, 36, 186

\bibitem[{Samolyk}(2009)]{2009JAVSO..37...44S}
{Samolyk}, G. 2009, JAAVSO, 37, 44

\bibitem[{Samolyk}(2010{\natexlab{a}})]{2010JAVSO..38..183S}
{Samolyk}, G. 2010{\natexlab{a}}, JAAVSO, 38, 183

\bibitem[{Samolyk}(2010{\natexlab{b}})]{2010JAVSO..38...85S}
{Samolyk}, G. 2010{\natexlab{b}}, JAAVSO, 38, 85

\bibitem[{Samolyk}(2011{\natexlab{a}})]{2011JAVSO..39...94S}
{Samolyk}, G. 2011{\natexlab{a}}, JAAVSO, 39, 94

\bibitem[{Samolyk}(2011{\natexlab{b}})]{2011JAVSO..39..177S}
{Samolyk}, G. 2011{\natexlab{b}}, JAAVSO, 39, 177

\bibitem[{Samolyk}(2014)]{2014JAVSO..42..426S}
{Samolyk}, G. 2014, JAAVSO, 42, 426

\bibitem[{Shapley}(1948)]{1948HarMo...7..249S}
{Shapley}, H. 1948, Harvard Observatory Monographs, 7, 249

\bibitem[{van Hamme}(1993)]{1993AJ....106.2096V}
{van Hamme}, W. 1993, \aj, 106, 2096

\bibitem[{Vilhu}(1982)]{1982A&A...109...17V}
{Vilhu}, O. 1982, \aap, 109, 17

\bibitem[{Wadhwa}(2005)]{2005Ap&SS.300..289W}
{Wadhwa}, S.~S. 2005, \apss, 300, 289

\bibitem[{Webbink}(1976)]{1976ApJS...32..583W}
{Webbink}, R.~F. 1976, \apjs, 32, 583

\bibitem[{Wilson}(1979)]{1979ApJ...234.1054W}
{Wilson}, R.~E. 1979, \apj, 234, 1054

\bibitem[{Wilson}(1990)]{1990ApJ...356..613W}
{Wilson}, R.~E. 1990, \apj, 356, 613

\bibitem[{Wilson} \& {Devinney}(1971)]{1971ApJ...166..605W}
{Wilson}, R.~E., \& {Devinney}, E.~J. 1971, \apj, 166, 605

\bibitem[{Zhang} \& {Zhang}(2006)]{2006NewA...11..339Z}
{Zhang}, X.~B., \& {Zhang}, R.~X. 2006, \na, 11, 339

\end{thebibliography}

\label{lastpage}

\end{document}